\documentclass{aa}  
\usepackage{amsmath}
\usepackage{natbib,twoopt}
\usepackage{breqn}
\usepackage[breaklinks=true, colorlinks=true, allcolors=blue]{hyperref} 
\makeatletter
\newcommandtwoopt{\citeads}[3][][]{\href{http://adsabs.harvard.edu/abs/#3}%
{\def\hyper@linkstart##1##2{}%
\let\hyper@linkend\@empty\citealp[#1][#2]{#3}}}
\newcommandtwoopt{\citepads}[3][][]{\href{http://adsabs.harvard.edu/abs/#3}%
{\def\hyper@linkstart##1##2{}%
\let\hyper@linkend\@empty\citep[#1][#2]{#3}}}
\newcommandtwoopt{\citetads}[3][][]{\href{http://adsabs.harvard.edu/abs/#3}%
{\def\hyper@linkstart##1##2{}%
\let\hyper@linkend\@empty\citet[#1][#2]{#3}}}
\newcommandtwoopt{\citeyearads}[3][][]%
{\href{http://adsabs.harvard.edu/abs/#3}
{\def\hyper@linkstart##1##2{}%
\let\hyper@linkend\@empty\citeyear[#1][#2]{#3}}}
\makeatother

\usepackage{graphicx}
\usepackage{txfonts}
\begin{document}

   \title{The {\it NuSTAR} view of the changing look AGN ESO 323-G77}

   \author{Roberto Serafinelli
          \inst{1}
          \and
          Valentina Braito\inst{2,3}
          \and
          James N. Reeves\inst{3,2}
          \and
          Paola Severgnini\inst{2}
          \and 
          \\Alessandra De Rosa\inst{1}
          \and
          Roberto Della Ceca\inst{2}
          \and
          Tracey Jane Turner\inst{4}
          }

   \institute{INAF - Istituto di Astrofisica e Planetologia Spaziali, Via del Fosso del Cavaliere 100, 00133, Roma, Italy\\
              \email{roberto.serafinelli@inaf.it}
         \and
          INAF - Osservatorio Astronomico di Brera, Via Brera 28, 20121, Milano, Italy \& Via Bianchi 46, Merate (LC), Italy
          \and
          Department of Physics, Institute for Astrophysics and Computational Sciences, The Catholic University of America, Washington, DC, 20064, USA
          \and
Eureka Scientific, Inc, 2452 Delmer St. Suite 100, Oakland, CA 94602, USA
             }

   \date{Received XXX; accepted YYY}

 
  \abstract
   {The presence of an obscuring torus at parsec-scale distances from the central black hole is the main ingredient for the Unified Model of Active Galactic Nuclei (AGN), as obscured sources are thought to be seen through this structure. However, the Unified Model fails to describe a class of sources that undergo dramatic spectral changes, transitioning from obscured to unobscured and vice-versa through time. The variability in such sources, so-called Changing Look AGN (CLAGN), is thought to be produced by a clumpy medium at much smaller distances than the conventional obscuring torus. ESO 323-G77 is a CLAGN that was observed in various states through the years with {\it Chandra}, {\it Suzaku}, {\it Swift}-XRT and {\it XMM-Newton}, from unobscured ($N_{\rm H}<3\times10^{22}$ cm$^{-2}$) to Compton-thin ($N_{\rm H}\sim1-6\times10^{23}$ cm$^{-2}$) and even Compton-thick ($N_{\rm H}>1\times10^{24}$ cm$^{-2}$), with timescales as short as one month. We present the analysis of the first {\it NuSTAR} monitoring of ESO 323-G77, consisting of 5 observations taken at different timescales (1, 2, 4 and 8 weeks from the first one) in 2016-2017, in which the AGN was caught in a persistent Compton-thin obscured state ($N_{\rm H}\sim2-4\times10^{23}$ cm$^{-2}$). We find that a Compton-thick reflector is present ($N_{\rm H,refl}=5\times10^{24}$ cm$^{-2}$), most likely associated with the presence of the putative torus. Two ionized absorbers are unequivocally present, located within maximum radii of $r_{\rm max,1}=1.5$ pc and $r_{\max,2}=0.01$ pc. In one of the observations, the inner ionized absorber is blueshifted, indicating the presence of a possible faster ($v_{\rm out}=0.2c$) ionized absorber, marginally detected at $3\sigma$. Finally, we are able to constrain the coronal temperature and the optical depth of ESO 323-G77, obtaining $kT_e=38$ keV or $kT_e=36$ keV, and $\tau=1.4$ or $\tau=2.8$, depending on the coronal geometry assumed.}


   \maketitle
%

\section{Introduction}
\label{sec:intro}

Observations of active galactic nuclei (AGN) reveal the presence of two main classes of sources. Type 1 AGN are sources for which the optical spectra show both narrow (FWHM$\leq1000$ km s$^{-1}$) and broad (FWHM$>1000$ km s$^{-1}$) lines, while type 2 AGN are objects whose spectra only manifest narrow lines. This suggests that in type 1 AGN the broad line region (BLR) is visible, while type 2 AGN have the BLR covered by obscuring material. The dichotomy between type 1 and type 2 objects led to a unification scheme based on the orientation of the AGN \citep[e.g.,][]{antonucci93,urry95}, where the central engine is surrounded by an axisymmetric absorber, called the torus, and the amount of obscuration is entirely due to the line of sight angle with respect to the AGN axis.\\ 
\indent According to the unification model, the column density $N_{\rm H}$ measured in X-ray spectra should follow this simple physical scheme. However, in many AGN the amount of obscuration in the X-rays is variable on a wide range of timescales \citep[e.g.,][]{risaliti02,markowitz14,laha20}, suggesting that the unification model is too simplistic to properly describe the whole phenomenon in detail. In particular, in some cases the X-ray absorbing medium is variable on very short timescales (days/weeks), which implies that the obscuring medium is clumpy and located at much smaller distances than the torus, possibly consistent with the BLR \citep[e.g.,][]{risaliti07,bianchi09,maiolino10,sanfrutos13,marinucci13,walton14}. In other cases, the X-ray absorption variability timescale is of the order of months or years \citep[e.g.,][]{piconcelli07,rivers11,coffey14,rivers15,ricci16,pizzetti22}, suggesting an origin from the putative circumnuclear torus. However, these results strongly depend on the observation sampling time; frequently adopted monthly observational monitoring may lose the variations at lower timescales. These findings suggest that the X-ray obscurer is not a single homogeneous entity, but rather the observational product of multiple layers of absorbing material from the BLR and the torus.\\ 
\indent Moreover, there is mounting evidence for a clumpiness of the circumnuclear torus \citep[e.g.,][]{tristram07}, which would imply that the probability of observing the central engine is always non-zero \citep[e.g.,][]{elitzur08,elitzur12}. The X-ray obscuration can therefore occur due to individual clumps passing through the line of sight, either in the BLR or in the circumnuclear torus.\\
\indent ESO 323-G77 is a nearby Seyfert 1 galaxy at redshift $z=0.015$, with a complex and highly variable absorber. A $\sim20$ ks observation by {\it XMM-Newton} in 2006 unveiled complex absorpion and emission features that revealed the presence of outflowing material \citep{jimenez-bailon08}. Subsequent observations with {\it XMM-Newton} (2013), {\it Chandra} (2011), {\it Swift}-XRT (2006) and {\it Suzaku} (2011) revealed a wide range of spectral shapes, mainly driven by variations of the column density of a neutral absorber at several timescales \citep{miniutti14,sanfrutos16}.\\
\indent The spectral shape of the source ranges from an unobscured state ($N_{\rm H}<10^{22}$ cm$^{-2}$) in four {\it Chandra} observations taken in 2010, a moderately absorbed state ($N_{\rm H}\sim3\times10^{22}-10^{23}$ cm$^{-2}$) for the 2006 {\it XMM-Newton} observation and two 2006 {\rm Swift}-XRT snapshots, a Compton-thin obscured state ($N_{\rm H}\sim1-6\times10^{23}$ cm$^{-2}$) observed by {\it XMM-Newton} in 2013 and in one {\it Swift}-XRT pointing in 2006, and finally a Compton-thick obscured state ($N_{\rm H}>10^{24}$ cm$^{-2}$) in the {\it Suzaku} observation taken in 2011. \cite{miniutti14} argued that low column density states ($N_{\rm H}\lesssim10^{23}$ cm$^{-2}$) are due to the presence of a clumpy obscuring torus, while the states with larger column densities are produced by obscuration by clumps of a closer medium, likely co-spatial with the BLR. This is reminiscent of other changing look sources such as NGC 1365 \citep[e.g.,][]{risaliti07}.\\
\indent Here we report the spectral analysis of the first {\it Nuclear Spectroscopic Telescope Array} \citep[NuSTAR,][]{harrison13} observations of ESO 323-G77. The paper is organized as follows. In Sect.~\ref{sec:data} we describe the data used for this work and the data reduction pipeline. Sect.~\ref{sec:spectral} describes the spectral analysis and all the models tested for the data. In Sect.~\ref{sec:discussion} we discuss the spectral models adopted, and in Sect.~\ref{sec:concl} we summarize our results.\\ 
\indent We adopt a standard flat cosmology with $H_0=70$ km s$^{-1}$ Mpc$^{-1}$, $\Omega_{\rm m}=0.3$ and $\Omega_\Lambda=0.7$.
\section{Data reduction}
\label{sec:data}

\begin{table}
\centering
\caption{The {\it NuSTAR} observations considered in this work. The exposures listed here are to be read as net exposures per single FPM module.}
\label{tab:data}
\begin{tabular}{lccc}
\hline
Epoch & OBSID & Date & Exposure (s)\\
\hline
1 & 60202021002 & 2016-12-14 & 39360\\
2 & 60202021004 & 2016-12-20 & 42531\\
3 & 60202021006 & 2017-01-04 & 43403\\
4 & 60202021008 & 2017-02-03 & 43295\\
5 & 60202021010 & 2017-03-31 & 38231\\
\hline
\end{tabular}
\end{table}

\begin{figure}
\centering
\includegraphics[scale=0.35]{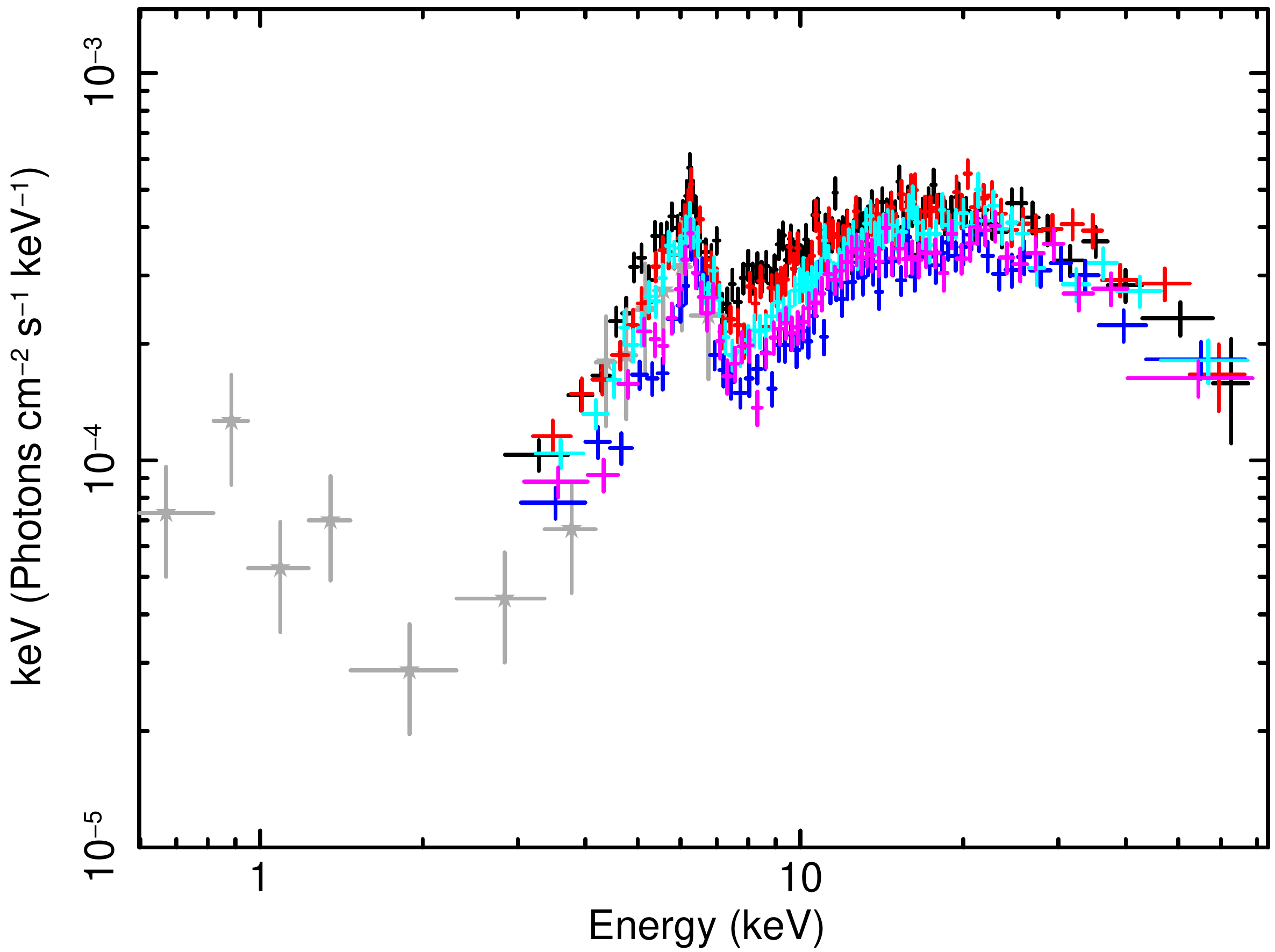}
\caption{Unfolded spectrum of the data analyzed here, adopting a simple model with an absorbed continuum powerlaw with $\Gamma=2$. Red, blue, cyan, black and magenta spectra mark the {\it NuSTAR} spectra of Epochs 1 to 5, respectively. The grey spectrum is the {\it Swift}-XRT one.}
\label{fig:dataraw}
\end{figure}

We analyze here a campaign of five {\it NuSTAR} observations performed between December 2016 and March 2017 for a total of $\sim200$ ks. Each observation has an exposure of approximately $40$ ks, taken at 1,2,4 and 8 weeks from the first one (see Table~\ref{tab:data} for details). The observations were coordinated with $\sim2$ ks of {\it Neil Gehrels Swift Observatory} \citep{gehrels04} snapshots taken with the X-ray Telescope (XRT). The {\it NuSTAR} spectra were reduced using the standard {\tiny HEASOFT} v6.28 command {\tiny NUPIPELINE} from the {\tiny NUSTARDAS} software package, using the most recent {\tiny CALDB} version. We filtered passages through the South Atlantic Anomaly by setting the task {\tiny NUCALSAA} 'optimized' mode. The two Focal Plane Module (FPM) source spectra A and B were extracted from a circular region with a radius of $40"$, centered on the source, while the background spectra were extracted from two circular regions with a radius of $45"$ each, on the same chip. The two FPMA and FPMB spectra were combined and the resulting spectrum was binned to a minimum of $50$ counts per bin. The energy band considered for our fits is in the range $E=3-65$ keV. The spectra of the {\it Swift}-XRT observations were extracted with the {\tiny HEASOFT} command {\tiny XSELECT}, selecting a circular region with a $30"$ radius. Background spectra were also extracted with the same procedure, but selecting a source-free circular region of $70"$ radius. The {\tiny XRTMKARF} task was used to produce ancillary files, and the response was provided by the {\tiny CALDB} repository. Given the negligible variability, the XRT spectra were all combined together, and grouped at a minimum of 10 counts per energy bin. The energy range $0.5-10$ keV was considered for the spectral fits. The folded spectra, adopting a simple powerlaw with photon index $\Gamma=2$, are shown in Fig.~\ref{fig:dataraw}.

\section{Spectral analysis}
\label{sec:spectral}

All spectral fits are performed using the software {\footnotesize XSPEC} v12.12.0 \citep{arnaud96}. We adopt a constant Galactic absorption described by a column density of $N_{\rm H}=7.75\times10^{20}$ cm$^{-2}$ \citep{HI4PI16}, which is modelled with {\tiny TBABS}, in all our models. In all models, a cross-correlation constant between XRT and {\it NuSTAR} ($C_{XRT/NuSTAR}$) is adopted. In every model the best-fit for this constant is $C_{XRT/NuSTAR}=0.8\pm0.1$. All errors on the best-fit parameters are given with a $90\%$ confidence level, i.e. $\Delta\chi^2=2.71$.

\subsection{Slab-reflection model}
We first tested a simple absorbed continuum powerlaw plus a scattered powerlaw, with tied photon index. Fe K$\alpha$ at $E=6.4$ keV, and Fe K$\beta$ at $E=7.06$ keV emission lines are also included, with fixed centroid energies and width ($\sigma=0.03$ keV). The Fe K$\beta$ line normalization is fixed at $13\%$ of the K$\alpha$ line \citep[e.g.,][]{palmeri03}. However, this simple model does not properly describe the current data set. As a very flat photon index ($\Gamma\sim1.35$) and an unacceptable statistic ($\chi^2/{\rm dof}=1751/907$) are obtained, it is clear that this model does not properly fit the data. Moreover, an equivalent width of EW$>250$ eV is obtained for the Fe K$\alpha$ line, which is a signature of the presence of a reflection component in obscured sources \citep[e.g.,][]{krolik94}. Therefore, we test a model that includes an absorbed powerlaw and a neutral reflector. The slab-reflection model {\tiny PEXRAV} \citep{magdziarz95} is used with Fe K$\alpha$ and Fe K$\beta$ emission lines modelled by two {\tiny ZGAUSS} components, plus an absorbed main powerlaw {\tiny ZPHABS*CABS*ZPOW}, and a soft-scattered powerlaw {\tiny ZPOW}. The overall model is 

\begin{dmath}
{\tt TBabs*(const1*cabs*zphabs*zpow1}+{\tt pexrav+zgauss1+ zgauss2+const2*zpow2)}.
\end{dmath}

\noindent The five NuSTAR spectra are fitted together, keeping all parameters tied among different epochs to the ones of Epoch 4, which is the brightest observation, with the exception of the column density $N_{\rm H}$ of the absorber and the normalizations of the main powerlaw and the reflection component, to take their variability into account. The {\it Swift}-XRT spectrum has all parameters tied to Epoch 4, as in all models considered in this work. We assume that the Fe K lines do not vary, as in most absorbed AGN \citep[e.g.,][]{fukazawa16}, and we keep the Fe K$\beta$ normalization fixed at $13\%$ of the value of the Fe K$\alpha$. All parameters of the scattered powerlaw are kept tied to the ones of the main one, whereas the two constants, {\tiny CONST1} kept fixed at 1 at all epochs, while {\tiny CONST2} is fitted for Epoch 4 and not allowed to vary, in order to take their ratio into account.\\
\indent The continuum is characterized by a photon index $\Gamma=1.75\pm0.03$ and a normalization that varies from $n_{\rm pl}=9^{+5}_{-4}\times10^{-4}$ to $n_{\rm pl}=(2.0\pm0.3)\times10^{-3}$ photons cm$^{-2}$ s$^{-1}$ keV$^{-1}$, while the second constant is kept free and is $\sim10^{-2}$. {\tiny PEXRAV} models a pure reflection component from an infinite slab, meaning that the reflection constant is fixed to $\mathcal{R}=-1$. The photon index of the reflection component is tied to the one of the main continuum, while the cut-off energy is fixed to $E_{\rm cut}=500$ keV. The line of sight absorption is given by $N_{\rm H}\sim(7-11)\times10^{23}$ cm$^{-2}$, depending on the observation, and therefore the model cabs is included to take into account the suppression of the continuum due to electron scattering, which is non-negligible at column densities larger than $5\times10^{23}$ cm$^{-2}$ \citep[e.g.,][]{yaqoob12}, with column density fixed to the value of {\tiny ZPHABS}.\\
\indent The Fe K$\alpha$ emission line centroid is found at $E=6.28^{+0.06}_{-0.07}$ keV, with width $\sigma=0.2\pm0.1$ keV and normalization $n_{{\rm FeK}\alpha}=(1.3\pm0.4)\times10^{-5}$ photons cm$^{-2}$ s$^{-1}$ keV$^{-1}$. The range of the equivalent width of the Fe K$\alpha$ emission line is EW$\sim0.2-0.4$ keV. The cut-off energy in the reflection spectrum is fixed at $E_{\rm cut}=500$ keV, while the photon index is tied to that of the continuum component. The abundances are fixed to solar ones, and the reflector normalizations vary in the range $n_{\rm refl}\sim(4-5)\times10^{-3}$ photons cm$^{-2}$ s$^{-1}$ keV$^{-1}$.\\
\indent The model has an overall goodness of fit of $\chi^{2}/{\rm dof}=978/909=1.07$.\\

\subsection{Toroidal model MYTORUS}
\label{sec:myt}

\begin{figure}
\centering
\includegraphics[scale=0.35]{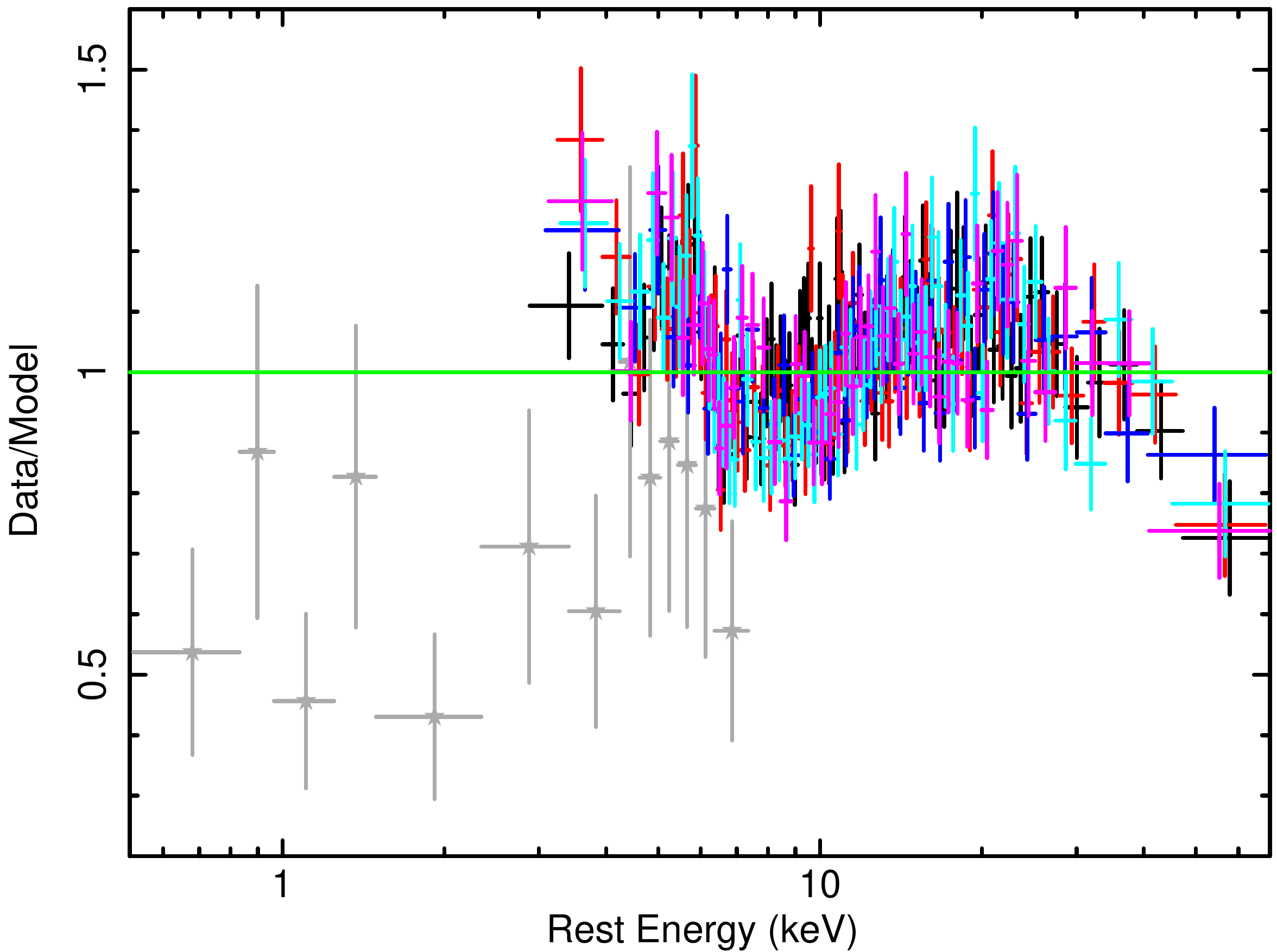}
\caption{Data-to-model ratio for the model in Eq.~\ref{eq:myt_noionized}, where the reflector is modelled with {\tiny MYTS}$_0$, and only the neutral absorber is considered. There are still significant ratios in the whole analyzed band, in particular the curvature is not well modelled by a single neutral absorber.}
\label{fig:noion}
\end{figure}

The disk-reflection model provides an acceptable goodness of fit. However, as pointed out by \citet{yaqoob12}, the model is inadequate to describe the reflector in detail. Indeed, the reflection spectrum assumes an infinite line-of-sight column density and it does not consider the finite nature of the reflector, as it was created assuming a point source illuminating an infinite slab.\\
\indent Therefore, in the following we adopt a detailed toroidal reflection model, {\tiny MYTORUS} \citep{murphy09}. This model assumes a toroidal geometry characterized by a column density $N_{\rm H}$ and a fixed covering factor of $0.5$, corresponding to a torus opening angle of $60^\circ$. Since the column density of ESO 323-G77 is variable, we adopt the decoupled standard model, in which the column density of the absorber $N_{\rm H,abs}$ is different from the column density of the reflector $N_{\rm H,refl}$ \citep{yaqoob12}. As a first step we multiply the continuum power law by the zeroth-order component of the model, namely the {\tiny XSPEC} table MYTZ\footnote{All MYTORUS tables are available at \url{http://mytorus.com/model-files-mytorus-downloads.html}. The MYTZ model can be downloaded with the table \texttt{mytorus\_Ezero\_v00.fits}}. This table allows us to evaluate the line-of-sight column density $N_{\rm H}$ of the absorber. We consider the angle $\theta$, which is the inclination angle between the polar axis of the absorber and the line of sight. For the MYTZ, we fix $\theta=90^\circ$, which corresponds to a line of sight direction for the absorber. We model the Compton hump continuum due to neutral reflection with the additive table MyTS$_0$\footnote{File name {\tt mytorus\_scatteredH200\_v00.fits}}. We fix $\theta=0^\circ$ to assume that this reflected component does not come from the line of sight. The column density of this component is independent from the line of sight one (decoupled model), and the normalization and photon index are kept fixed to the ones of the continuum. The Fe K$\alpha$ and Fe K$\beta$ emission lines of the line-of-sight reflection are included with the additive table MyTL$_0$\footnote{File name {\tt mytl\_V000010nEp000H200\_v00.fits}}, with fixed value $\theta=0^\circ$, normalization and $\Gamma$ tied to the absorbed continuum values. We multiply MyTL$_0$ by the convolution model {\tiny GSMOOTH}, to take into account the broadening of the iron line. We fix the line width in the model to $\sigma=0.03$ keV, following the upper limit found by \cite{sanfrutos16} with {\it Chandra} HETG. The fit has a global statistic of $\chi^2/{\rm dof}=1129/915=1.23$.\\
\indent We also allow for a forward scattering component on the line of sight, namely another Compton-reflected continuum with fixed $\theta=90^\circ$ (hereafter MyTS$_{90}$). We assume that the column density ($N_{{\rm H},0}$) of this component coincides with the line-of-sight $N_{\rm H}$. This additional reflection component is also accompanied by a table with iron lines MyTL$_{90}$, where the column density is tied to $N_{{\rm H},0}$. MyTL$_{90}$ is also multiplied by a {\tiny GSMOOTH} model with fixed $\sigma=0.03$ keV. The normalizations and photon indices of MyTS$_{90}$ and MyTL$_{90}$ are also tied to the one of the main powerlaw. The goodness of fit is given by $\chi^2/{\rm dof}=1162/915=1.25$. This means that the reflection due to the absorbing material on the line of sight is not required in our model. In all models from here on, we will only consider the reflection component out of the line of sight.\\
\indent The model is therefore
\begin{dmath}
{\tt TBabs*((const1*MyTZ*zpow1}+{\tt MyTS_0+gsmooth*MyTL_0)}+{\tt const2*zpow2)}
\label{eq:myt_noionized}
\end{dmath}
\indent We find a line-of-sight $N_{\rm H, abs}$ ranging from $(3.2\pm0.3)\times10^{23}$ cm$^{-2}$ in Epoch 4 up to $(5.5\pm0.5)\times10^{23}$ cm$^{-2}$ at Epoch 2. The out of line of sight column density of the reflector is $N_{\rm H,refl}=4.0^{+0.3}_{-0.7}\times10^{24}$ cm$^{-2}$. We also obtain a flatter photon index $\Gamma=1.61\pm0.3$, with respect to the one obtained with the slab-reflection model.

\subsection{Ionized absorbers}

\begin{figure}
\centering
\includegraphics[scale=0.35]{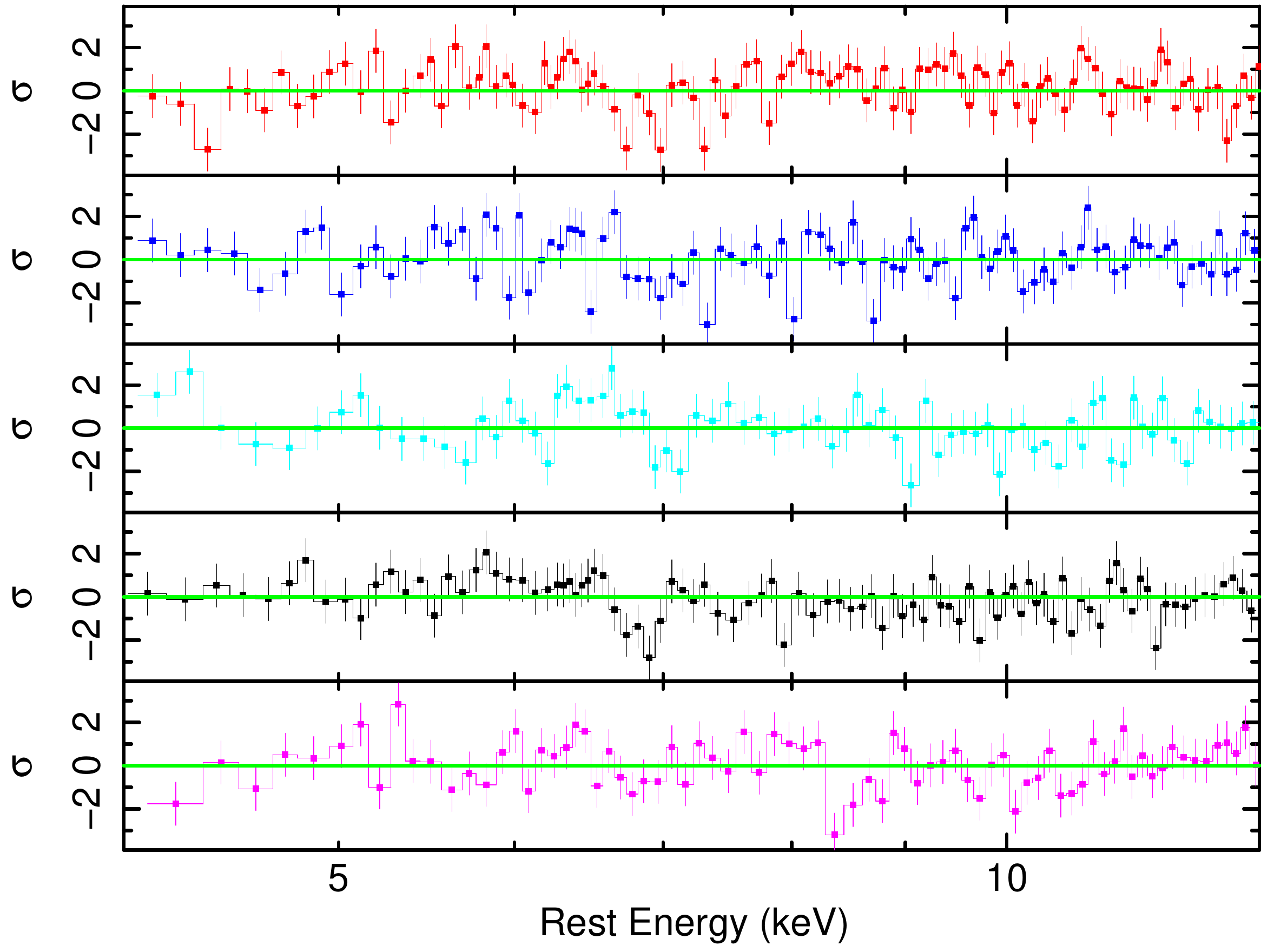}
\caption{Residuals when the model with only one ionized absorber is fitted. The observations are ordered top to bottom from the first to the last taken. An absorption complex at $\sim7$ keV is observed in Epochs 1, 3 and 4. At Epoch 5, the absorption complex is observed at $\sim8.5$ keV, suggesting a possible outflowing velocity of $v\sim0.2c$.}
\label{fig:resi_1abs}
\end{figure}

\begin{figure}
\centering
\includegraphics[scale=0.35]{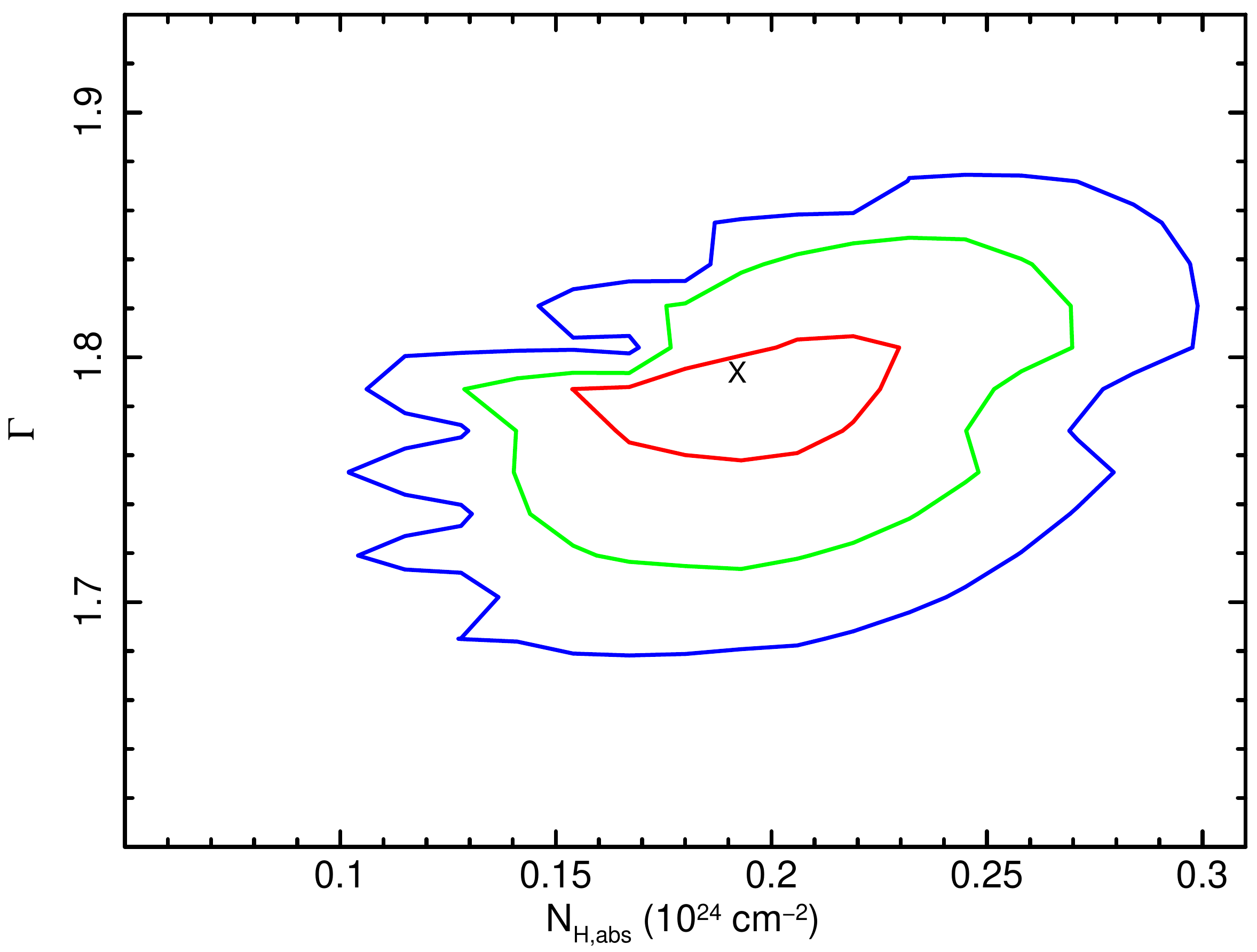}
\caption{Contour plot of the spectral slope $\Gamma$ and the line of sight column density $N_{\rm H}$ obtained for Epoch 4, as obtained from the zero-order {\tiny MYTORUS} model. The red, green and blue line represent $68\%$ ($1\sigma$), $95\%$ ($2\sigma$) and $99.7\%$ ($3\sigma$) contours.}
\label{fig:contgammanh}
\end{figure}

\begin{figure}
\centering\includegraphics[scale=0.35]{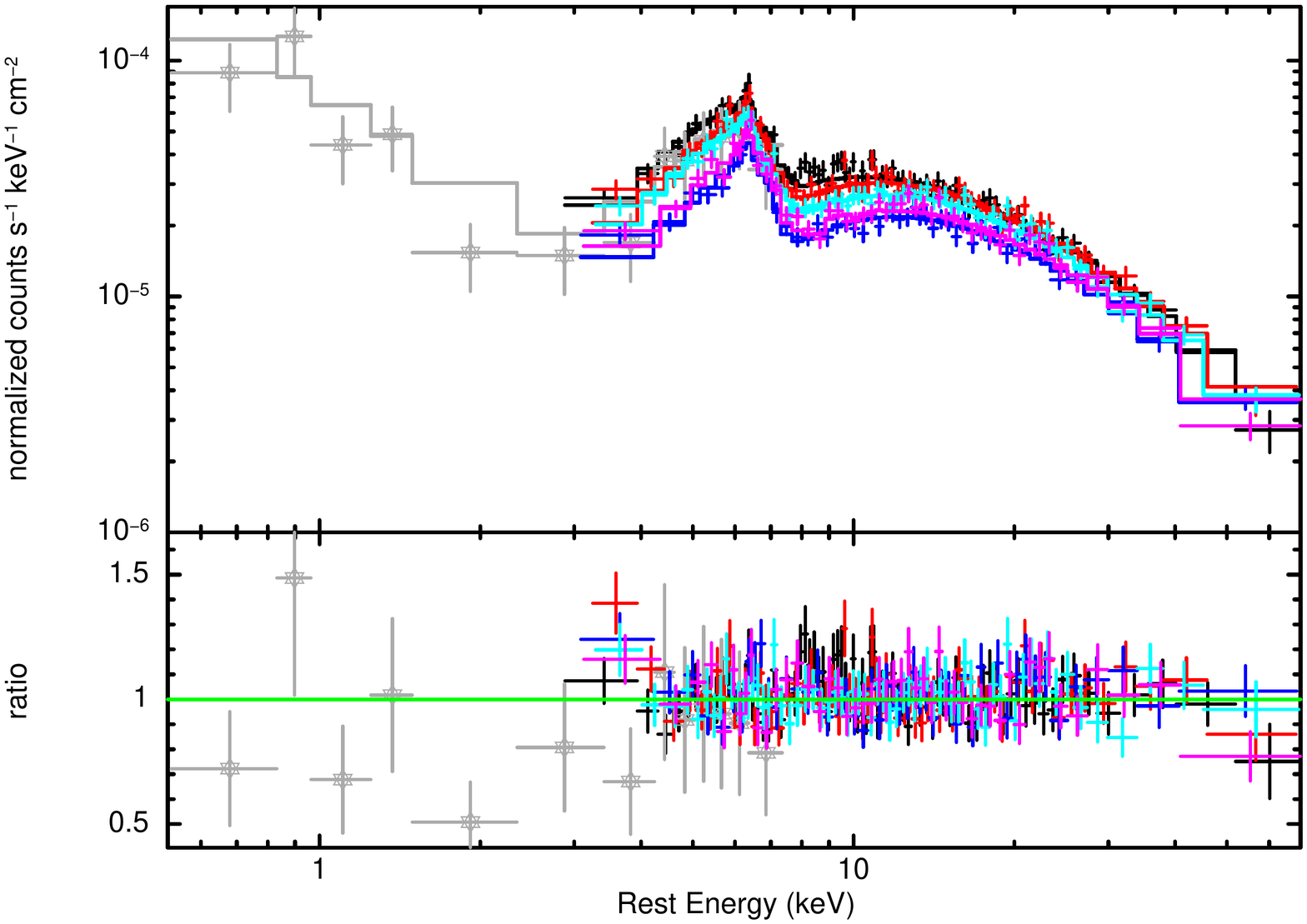}
\caption{Normalized spectra and data-to-model ratio of ESO 323-G77. The {\footnotesize MYTORUS} model is shown here. The same color code used in Fig.~\ref{fig:resi_1abs} is adopted, with the addition of the {\it Swift}-XRT spectrum, shown in grey.}
\label{fig:mytdata}
\end{figure}

\begin{figure}
\centering
\includegraphics[scale=0.35]{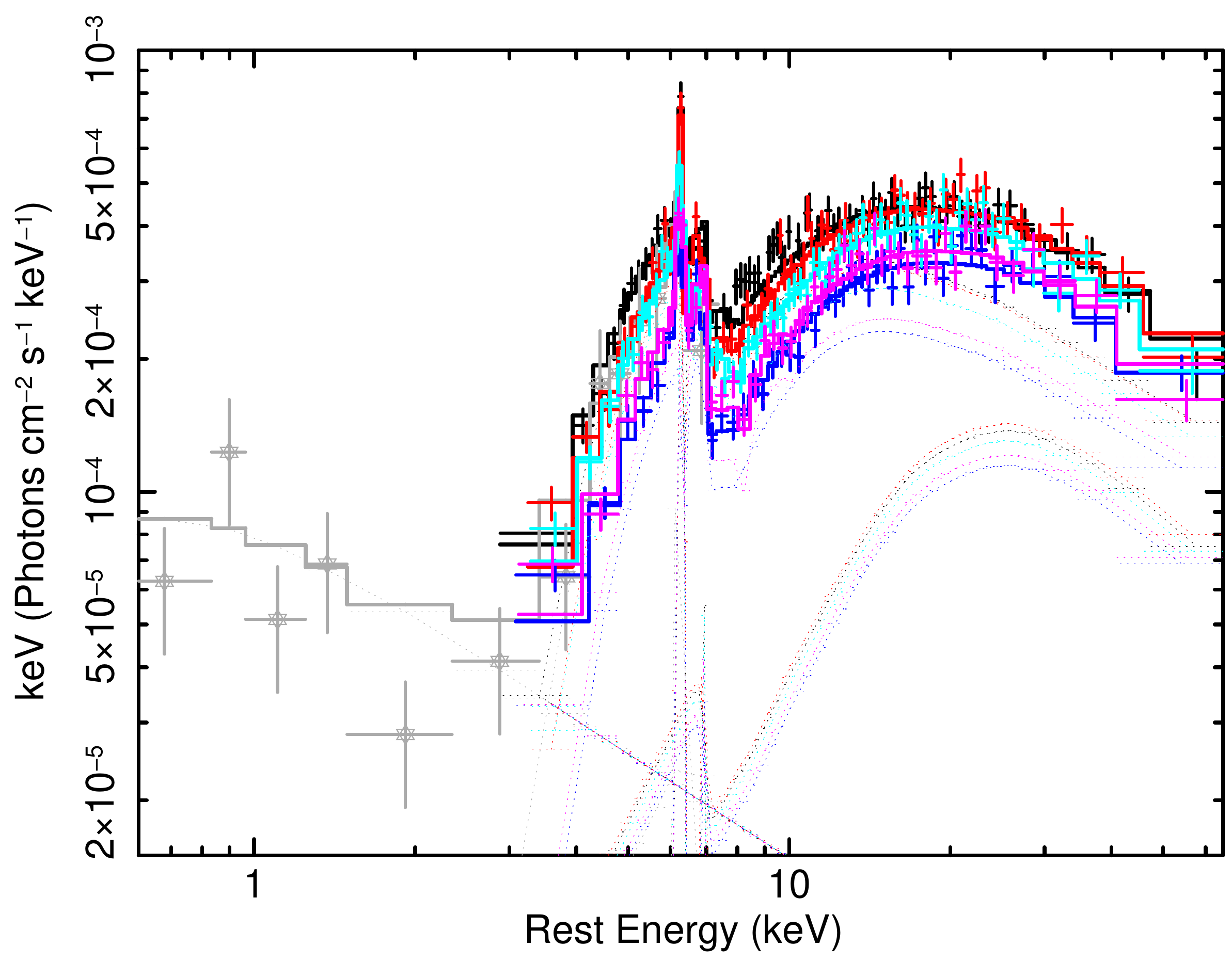}
\caption{Unfolded spectra of ESO 323-G77, determined from the {\it Swift}-XRT and NuSTAR data, based on the {\footnotesize MYTORUS} model.}
\label{fig:myteuf}
\end{figure}

\begin{table*}
\centering
\caption{Best-fit parameters of the final {\tiny MYTORUS} model shown in Eq.~\ref{eq:2ion_mytfinal}. The goodness of fit is $\chi^2/{\rm dof}=927/906$. The {\tiny COMPTT} parameters are taken from the best-fit model shown in Eq.~\ref{eq:mytcomptt}, assuming a comptonized continuum produced by a corona with a slab geometry.}
\label{tab:myttab}
\begin{tabular}{lccccccccc}
\hline
Parameter & Epoch 1 & Epoch 2 & Epoch 3 & Epoch 4 & Epoch 5\\
\hline
\\
Central source ({\tt zpow})\\
$\Gamma$ & $-$ & $-$ & $-$ & $1.79^{+0.04}_{-0.06}$ & $-$\\
\\
${\rm norm}$ $(10^{-3}$ photons cm$^{-2}$ s$^{-1}$ keV$^{-1})$ & $4.2\pm0.7$ & $3.4\pm0.6$ & $3.9\pm0.7$ & $4.0\pm0.7$ & $3.5\pm0.6$\\
\\
$F^{\rm scat}/F^{\rm nucl}$ & $-$ & $-$ & $-$ & $2.3^{+0.4}_{-0.3}\times10^{-2}$ & $-$\\
\\
\hline
\\
Central source ({\tt compTT}, slab corona)\\
$kT$ (keV) & $-$ & $-$ & $-$ & $38\pm2$ & $-$\\
\\
$\tau$ & $-$ & $-$ & $-$ & $1.4\pm0.1$ & $-$\\
\\
\hline
\\
Neutral absorber ({\tt MYTORUS})\\
MyTZ\\ 
$N_{\rm H,abs}$ ($10^{23}$ cm$^{-2}$) & $2.6^{+0.4}_{-0.5}$ & $3.7^{+0.6}_{-0.5}$ & $2.5\pm0.5$ & $1.9\pm0.4$ & $3.4^{+0.6}_{-0.3}$\\
\\
Reflection ({\tt MYTORUS})\\
MyTS$_{0}$\\
$N_{\rm H,refl}$ ($10^{24}$ cm$^{-2}$) & $-$ & $-$ & $-$ & $5.0^{+2.8}_{-1.3}$ & $-$\\
\\
\hline
\\
Ionized absorbers ({\tt xstar})\\
Zone 1 (external)\\
$N_{\rm H,z1}$ ($10^{23}$ cm$^{-2}$) & $-$ & $-$ & $-$ & $3.5^{+0.6}_{-0.7}$ & $-$\\
$\log\xi$ (erg cm s$^{-1}$) & $-$ & $-$ & $-$ & $2.4\pm0.1$ & $-$\\
\\
Zone 2 (internal)\\
$N_{\rm H,z2}$ ($10^{23}$ cm$^{-2}$) & $2^{+7}_{-1}$ & $<14$ & $6^{+16}_{-4}$ & $2^{+4}_{-1}$ & $2^{+7}_{-1}$\\
$\log\xi$ (erg cm s$^{-1}$) & $-$ & $-$ & $-$ & $4.0^{+0.5}_{-0.2}$ & $-$\\
$v/c$ & $0$ & $0$ & $0$ & $0$ & $0.21^{+0.02}_{-0.03}$\\
\\
\hline
\\
$C_{XRT/NuSTAR}$ & 0 & 0 & 0 & $0.8\pm0.1$ & 0\\
\\
Observed fluxes\\
$F^{\rm obs}_{2-10\;{\rm keV}}$ (erg cm$^{-2}$ s$^{-1}$) & $2.7\times10^{-12}$ & $1.8\times10^{-12}$ & $2.5\times10^{-12}$ & $3.1\times10^{-12}$ & $2.0\times10^{-12}$\\
\\
Unabsorbed fluxes\\
$F^{\rm unabs}_{2-10\;{\rm keV}}$ (erg cm$^{-2}$ s$^{-1}$) & $4.6\times10^{-12}$ & $3.0\times10^{-12}$ & $4.4\times10^{-12}$ & $5.5\times10^{-12}$ & $4.7\times10^{-12}$\\
\\
Reflection flux\\
$F^{\rm refl}_{3-65\;{\rm keV}}$ (erg cm$^{-2}$ s$^{-1}$) & $-$ & $-$ & $-$ & $9.5\times10^{-12}$ & $-$\\
\hline
\end{tabular}
\end{table*}

Fig.~\ref{fig:noion} shows the residuals of the model in Eq.~\ref{eq:myt_noionized}. There are significant residuals in the $E\sim5-10$ keV energy range and above $E\sim20$ keV, showing that it does not properly fit the curvature of the spectrum, which means that additional components might be needed. Since past observations of this source reported the presence of ionized absorbers \citep{jimenez-bailon08,miniutti14,sanfrutos16}, we consider the addition of one of such features. We denote this absorber as Zone 1. We adopt a grid of photoionized absorbers, produced with the {\tiny XSTAR} \citep{kallman01} photoionization code. The grid spans a relatively wide ionization ($\log(\xi/{\rm erg\;cm\;s^{-1}})\sim2-6$) and column density ($N_{\rm H}\sim5\times10^{22}-5\times10^{24}$ cm$^{-2}$) range. The turbulent velocity adopted to generate the grid is $v_{\rm turb}=3000$ km s$^{-1}$. We first allow the ionization and the column density to vary among different observations, but we do not find significant changes in either parameters, therefore we fix both parameters to the ones of Epoch 4. The addition of this component improves the fit by $\Delta\chi^2/\Delta{\rm dof}=159/2$, with the overall goodness of fit being $\chi^2/{\rm dof}=970/913$. The photon index is $1.74\pm0.06$. The column density of this absorber is given by $N_{\rm H,z1}=(5.8\pm0.5)\times10^{23}$ cm$^{-2}$ and the ionization is $\log\xi_{\rm z1}/({\rm erg\;cm\;s^{-1}})=2.6\pm0.1$.\\
\indent The addition of the Zone 1 absorber significantly reduces the curvature residuals in Fig.~\ref{fig:noion}. The residuals with the new model in the $E=4-13$ keV band are shown in Fig.~\ref{fig:resi_1abs}, where the data still show significant residuals in the Fe K$\alpha$ spectral region ($E=6-10$ keV) in almost all observations. Most observations show an absorbing structure around $6.5-7$ keV, which may be due to absorbing material. This is particularly noticeable near 7 keV in epochs 3 and 4 (see Fig.~\ref{fig:resi_1abs}, panels 3 and 4). Moreover, a second more ionized absorber was reported in \cite{jimenez-bailon08}, \cite{miniutti14} and \cite{sanfrutos16}, which could be responsible for this absorbing feature. We thus add a second absorber, which we label as Zone 2, using the same {\tiny XSTAR} grid used for the first one. We initally assumed that also this more ionized absorber did not vary between the 5 epochs. The addition of this absorber improves the statistic by $\Delta\chi^2/\Delta{\rm dof}=26/2$ to $\chi^2/{\rm dof}=944/911$. We obtain a photon index of $\Gamma=1.81^{+0.06}_{-0.07}$, a column density of $N_{\rm H,z2}=1.6^{+14.5}_{-0.7}\times10^{23}$ cm$^{-2}$ and a ionization parameter of $\log\xi_{\rm z2}/({\rm erg\;cm\;s}^{-1})=4.0^{+1.1}_{-0.2}$. Given its higher ionization, we assume that Zone 2 is closer to the black hole with respect to Zone 1. Since $N_{\rm H}$ and $\log\xi$ are notoriously degenerate, we keep the ionization at all epochs fixed to the one of Epoch 4, while all column densities are allowed to vary independently. The goodness of fit slightly improves to $\chi^2/{\rm dof}=938/907$.\\ 
\indent Finally, as shown in Fig.~\ref{fig:resi_1abs}, the absorber in Epoch 5 appears as a blueshifted absorption line at $E_{\rm rest}\sim8.5$ keV, which is a clear signature of a non-zero velocity. Hence, we free the velocity of the Zone 2 absorber in Epoch 5, in order to take this blueshift into account. We obtain $z_{\rm obs}=-0.18\pm0.02$, which corresponds\footnote{The observed shift $z_{\rm obs}$ is related to the rest-frame blueshift $z_{\rm abs}$ by the relation $z_{\rm abs}=(1+z_{\rm obs})/(1+z_c)-1$, where $z_c=0.015$ is the cosmological redshift. The velocity of the absorber is given by $v/c=(z_{\rm abs}^2+2z_{\rm abs})/(z_{\rm abs}^2+2z_{\rm abs}+2)$.} to a velocity $v=(0.21^{+0.02}_{-0.03})c$. The goodness of fit further improves by $\Delta\chi^2/\Delta{\rm dof}=11/1$ to a final value of $\chi^2/{\rm dof}=927/906=1.02$. The column density of the absorber in Zone 2 is constrained in four out of five observations, ranging from $N_{\rm H,z2}=2^{+7}_{-1}\times10^{23}$ cm$^{-2}$ (Epoch 1) to $N_{\rm H,z2}=6^{+16}_{-4}\times10^{23}$ cm$^{-2}$ (Epoch 3). We note that in Epoch 2 we can place only an upper limit on the column density. Indeed, Epoch 2 does not show a clear absorption signature in Fig.~\ref{fig:resi_1abs} (blue curve). The ionization parameter is $\log\xi_{\rm z2}/{\rm erg\;cm\;s}^{-1}=4.0^{+0.5}_{-0.2}$.\\
\indent The final model is therefore 

\begin{dmath}
{\tt TBabs*((const1*xstar1*MyTZ*xstar2*zpow1}+{\tt MyTS_0+gsmooth*MyTL_0)}+{\tt const2*zpow2)}
\label{eq:2ion_mytfinal}
\end{dmath}

where {\tiny XSTAR1} and {\tiny XSTAR2} are the ionized absorbers in Zone 1 and Zone 2, respectively. The photon index of the spectrum, after the addition of these two ionized absorbers, is $\Gamma=1.79^{+0.04}_{-0.06}$. Fig.~\ref{fig:contgammanh} shows the contour plot of $\Gamma$ with the MyTZ column density $N_{\rm H}$ for the brightest observation of the campaign, i.e. Epoch 4. The contour plot shows that both the photon index $\Gamma$ and the absorbing column density $N_{\rm H,abs}$ are well constrained at $3\sigma$ confidence level. The best-fit parameters obtained with this model are summarized in Table~\ref{tab:myttab}. The normalized spectrum with data-to-model ratios and the unfolded spectrum are shown in Figs.~\ref{fig:mytdata} and \ref{fig:myteuf}.\\
\indent We also tested an alternative approach in which the absorber column density $N_{\rm H,abs}$ is kept tied among the observations, while the photon index $\Gamma$ is allowed to vary. Unsurprisingly, the column density is $N_{\rm H}=(2.6\pm0.5)\times10^{23}$ cm$^{-2}$, which is the mean value of the $N_{\rm H}$ found independently when the parameter is allowed to vary between observations. We find various values of the photon index, ranging from $\Gamma=1.51\pm0.06$ for Epoch 3 to $\Gamma=1.84^{+0.05}_{-0.04}$ for Epoch 4. However, we obtain a worse fit statistic of $\chi^2/{\rm dof}=983/906$, which means that an absorber variation is favored. Notably, the smaller photon indices are also the ones with greater absorption and viceversa, resulting in an apparent steeper when brighter effect. This effect is driven by the absorption variability, as the source has historically experienced in the past, and should not be confused with the continuum softer when brighter effect, driven by intrinsic $\Gamma$ variations \citep[e.g.,][]{sobolewska09,serafinelli17}.

\subsection{Alternative model for the reflector: BORUS}

\begin{table*}
\centering
\caption{Best-fit parameters of the final {\tiny BORUS} model shown in Eq.~\ref{eq:borus_final}. The goodness of fit is $\chi^2/{\rm dof}=923/901$. The {\tiny NTHCOMP} parameters are taken from the best-fit model shown in Eq.~\ref{eq:borus_nthcomp}, assuming a Comptonized continuum produced by a spherical corona.}
\label{tab:borustab}
\begin{tabular}{lccccccccc}
\hline
Parameter & Epoch 1 & Epoch 2 & Epoch 3 & Epoch 4 & Epoch 5\\
\hline
\\
Central source ({\tt zcutoffpl})\\
$\Gamma$ & $-$ & $-$ & $-$ & $1.79^{+0.04}_{-0.06}$ & $-$\\
\\
${\rm norm}$ $(10^{-3}$ photons cm$^{-2}$ s$^{-1}$ keV$^{-1})$ & $3.0^{+0.3}_{-0.5}$ & $2.4^{+0.3}_{-0.5}$ & $2.8^{+0.3}_{-0.5}$ & $3.0^{+0.3}_{-0.4}$ & $2.5^{+0.3}_{-0.4}$\\
\\
$F^{\rm scat}/F^{\rm nucl}$ & $-$ & $-$ & $-$ & $3.1^{+0.5}_{-0.4}\times10^{-2}$ & $-$\\
\\
\hline
\\
Central source ({\tt nthcomp})\\
\\
$\Gamma$ & $-$ & $-$ & $-$ & $1.73^{+0.01}_{-0.05}$ & $-$\\
\\
$kT$ (keV) & $-$ & $-$ & $-$ & $36^{+13}_{-8}$ & $-$\\
\\
\hline
\\
Neutral absorber\\
Absorption ({\tt zphabs})\\ 
$N_{\rm H}$ ($10^{23}$ cm$^{-2}$) & $2.7\pm0.2$ & $2.7\pm0.3$ & $2.8\pm0.2$ & $3.0\pm0.2$ & $3.4\pm0.3$\\
\\
Reflection ({\tt borus02})\\
$N_{\rm H}$ ($10^{24}$ cm$^{-2}$) & $-$ & $-$ & $-$ & $2.7^{+0.4}_{-0.8}$ & $-$\\
Covering factor ($C_f$) & $-$ & $-$ & $-$ & $0.90^{+0.02}_{-0.03}$ & $-$\\
\\
\hline
\\
Ionized absorbers ({\tt xstar})\\
Zone 1 (external)\\
$N_{\rm H}$ ($10^{23}$ cm$^{-2}$) & $-$ & $-$ & $-$ & $2.6^{+0.9}_{-0.8}$ & $-$\\
$\log\xi$ (erg cm s$^{-1}$) & $-$ & $-$ & $-$ & $2.37^{+0.05}_{-0.25}$ & $-$\\
\\
Zone 2 (internal)\\
$N_{\rm H}$ ($10^{23}$ cm$^{-2}$) & $<2.2$ & $<1.4$ & $4\pm2$ & $1.6^{+0.1}_{-0.8}$ & $2^{+2}_{-1}$\\
$\log\xi$ (erg cm s$^{-1}$) & $-$ & $-$ & $-$ & $4.0^{+0.2}_{-0.1}$ & $-$\\
$v/c$ & $0$ & $0$ & $0$ & $0$ & $0.21\pm0.02$\\
\\
\hline
\end{tabular}
\end{table*}

We also test for a spherical reprocessor, using the model BORUS \citep{balokovic18}. We consider a continuum described by a cut-off power law, {\tiny ZCUTOFFPL}, with a line of sight absorption modelled by {\tiny ZPHABS}, and reflector described by the table {\tiny BORUS02}\footnote{All BORUS tables can be downloaded from the website \url{https://sites.astro.caltech.edu/~mislavb/download}}. We also include the two ionized absorbers located in Zone 1 and Zone 2. Also in this model we allow the column density of the Zone 2 high-ionization absorber $N_{\rm H}$ to vary between observations, while we assume the ionization parameter to remain constant between the observations of the campaign. The model used is

\begin{dmath}
{\tt TBabs*((const1*xstar1*zphabs*xstar2*zcutoffpl1}+{\tt borus02)}+{\tt const2*zcutoffpl2).}
\label{eq:borus_final}
\end{dmath}

We obtain a photon index $\Gamma=1.79^{+0.04}_{-0.06}$, consistent with the value obtained with the MYTORUS model. As the cut-off energy $E_{\rm cut}$ is unconstrained, we fix it to a fiducial $E_{\rm cut}=500$ keV. The neutral column density varies from $N_{\rm H,abs}=(2.7\pm0.3)\times10^{23}$ cm$^{-2}$ (Epochs 1 and 2) to $(3.4\pm0.3)\times10^{23}$ cm$^{-2}$ (Epoch 5), roughly consistent with the ones found with the MYTORUS model. The column density of the reprocessor is $N_{\rm H}=2.7^{+0.4}_{-0.8}\times10^{24}$ cm$^{-2}$, which is consistent to the value found in the MYTORUS model. The covering factor of the reprocessor is given by $C_f=0.90^{+0.02}_{-0.03}$. Finally, we obtain consistent values for the column density and the ionization parameter of the ionized absorber in Zone 1. The ionization parameter of the absorber in Zone 2 is also consistent with the one obtained with the MYTORUS model. The column density of the absorber in Zone 2 is also consistent, althought with large uncertainties. The goodness of fit of this model is given by $\chi^2/{\rm dof}=923/901$.\\

\subsection{Comptonizing plasma continuum}
\label{sec:comptt}

\indent It is also interesting to investigate the coronal parameters of this source, as these are often elusive for obscured sources. Therefore, we investigate physical Comptonization models for the continuum with both the {\tiny MYTORUS} and {\tiny BORUS} models. Starting from the {\tiny MYTORUS} model in Eq.~\ref{eq:2ion_mytfinal}, We adopted the same configuration  and free parameters, but we replaced the power law continuum with {\tiny COMPTT} \citep{titarchuk94}. We also adopted the appropriate {\tiny MYTORUS} table, namely we adopt the tables MYTS$^{\rm TT}_0$\footnote{File name {\tt mytorus_scatteredkT034_v00.fits}} and MyTL$^{\rm TT}_0$\footnote{File name {\tt mytl_V000010nEp000kT034_v00.fits}}, and we use them the same way we used MYTS$_0$ and MYTL$_0$ in Sect.~\ref{sec:myt}. The model is then

\begin{dmath}
{\tt TBabs*((const1*xstar1*MyTZ*xstar2*compTT1}+{\tt MyTS^{TT}_0+gsmooth*MyTL^{TT}_0)}+{\tt const2*compTT2).}
\label{eq:mytcomptt}
\end{dmath}

We first explore the slab coronal geometry by fixing the value of the parameter {\tt approx} to $0.5$. We do not find significant differences in any other parameter obtained in the previous section. The coronal temperature with this fit is $kT=38\pm2$ keV, while the optical depth is $\tau=1.4\pm0.1$. The goodness of fit of this model is given by $\chi^2/{\rm dof}=920/906$. Typically, assuming a spherical geometry in {\tiny COMPTT}, the best-fit coronal parameters would be a similar temperature, but a larger optical depth \citep[e.g.,][]{tortosa18}. However, the MYTS$^{\rm TT}_{\theta}$ tables do not include larger values of $\tau$, and therefore it is not possible to explore the parameters of a spherical geometry.\\
\indent However, the spherical geometry might be explored within the {\tiny BORUS} model shown in Eq.~\ref{eq:borus_final}. {\tiny BORUS12} is produced with the thermal comptonization continuum model {\tiny NTHCOMP} \citep{magdziarz95}, which assumes a spherical geometry for the corona. Hence, we also use such model for the continuum, and the model is therefore:\\

\begin{dmath}
{\tt TBabs*((const1*xstar1*zphabs*xstar2*nthcomp1}+{\tt borus12)}+{\tt const2*nthcomp2).}
\label{eq:borus_nthcomp}
\end{dmath}

\noindent with a goodness of fit of $\chi^2/{\rm dof}=915/901$. We obtain $\Gamma=1.73^{+0.01}_{-0.05}$ and a coronal temperature of $kT=36^{+13}_{-8}$ keV. Remarkably, this value is consistent with the {\tiny COMPTT} temperature obtained assuming a slab geometry in the {\tiny MYTORUS} model, even adopting a different continuum model.

\subsection{Relativistic reflection}

The presence of a possible relativistic iron line in the X-ray spectra of this AGN was inferred by \cite{jimenez-bailon08} during an unabsorbed state. Therefore, we test the possibility that such component could also be detected in an absorbed state, and we add the relativistic reflection component {\tiny RELXILL} \citep{garcia14,dauser14} to the model in Eq.~\ref{eq:2ion_mytfinal}. The global fit improves by $\Delta\chi^2/\Delta{\rm dof}=38/7$. All parameters with the exception of the normalization are kept tied between observations. We assume a frozen cut-off energy $E_{\rm cut}=500$ keV, a disk external radius of $R_{\rm out}=400R_g$, where $R_g=GM/c^2$ is the gravitational radius, a $45^\circ$ inclination \citep{schmid03}, a solar iron abundance and an emissivity index of $-3$. The spin of the black hole is unconstrained, for which therefore we freeze $a=0$, and we obtain a disk internal radius $R_{\rm in}<12 R_g$, consistent with the findings of \cite{miniutti14}. The disk ionization parameter is $\log(\xi/{\rm erg\;cm\;s}^{-1})>3$. A steeper photon index $\Gamma=1.87^{+0.04}_{-0.09}$ is found, although consistent with the one found with the model in Eq.~\ref{eq:2ion_mytfinal} at $90\%$ confidence level. The normalization of the relativistic component is unconstrained in Epoch 4, ${\rm norm}_{\rm relx,4}<8\times10^{-6}$ photons cm$^{-2}$ s$^{-1}$ keV$^{-1}$, while in Epochs 2, 3 and 5 it is roughly constant (${\rm norm}_{\rm relx,2,3,5}=6^{+7}_{-4}\times10^{-6}$ photons cm$^{-2}$ s$^{-1}$ keV$^{-1}$), and in Epoch 1 it is ${\rm norm}_{\rm relx,1}=10^{+6}_{-5}\times10^{-6}$ photons cm$^{-2}$ s$^{-1}$ keV$^{-1}$. We do not find significant differences in the absorbing column density from Table~\ref{tab:myttab}. However, the two reflectors are degenerate, and therefore we find a lower limit for the neutral reflector column density, $N_{\rm H,refl}>4\times10^{24}$ cm$^{-2}$, even though it is consistent with the value of Table~\ref{tab:myttab}.\\ 
\indent We also tested {\tiny RELXILL} as an additional reflection component in the model  where we assume a comptonizing continuum {\tiny COMPTT} (Eq.~\ref{eq:mytcomptt}), to test the possible influence on the measure of $kT$ and $\tau$. The temperature of the corona is $kT=26\pm9$ keV, and $\tau=1.5^{+0.3}_{-0.1}$, consistent within the $3\sigma$ contour plot of these two parameters for the model without a disk reflection component (see Fig.~\ref{fig:contkTtau}). Very similar results are obtained by testing {\tiny RELXILL} on the two models that use {\tiny BORUS} for the neutral reflection.\\
\indent We stress that the {\tiny RELXILL} component contributes to $\lesssim10\%$ of the $2-10$ keV observed flux, and the main changes in this model are in the spectral region between $3$ and $5$ keV, where {\it NuSTAR} is less sensitive. Also, many parameters of the relativistic reflection model are unconstrained due to the complex model and numerous degeneracies with the neutral reflector. We point that, in order to accurately measure the parameters of the ionized relativistic reflection within the framework of such a complex spectral model, a broad band spectrum and an improved energy resolution would be needed. For instance, a simultaneous {\it XMM-Newton} and {\it NuSTAR} observation would be ideal to observe the Fe K$\alpha$ spectral region in detail.

\section{Discussion}
\label{sec:discussion}


\subsection{Comparison between MYTORUS and BORUS models}
The {\tiny MYTORUS} model has been built assuming a toroidal shape, asymmetric on the azimuthal axis. The covering factor in such model is kept fixed by assuming that the torus opening angle is $\theta_{\rm OA}=60^\circ$, which means that its value is $C_f=\cos(\theta_{\rm OA})=0.5$ \citep{murphy09}. Conversely, {\tiny BORUS} has a spherical geometry for the reprocessor, with polar cutouts corresponding to a variable opening angle $\theta_{\rm OA}$, and therefore is able to fit a value for the $C_f$, ranging from $C_f=0.1$ to $C_f=1$ \citep{balokovic18}. The two best-fit values of the average column density of the reflector are slightly different, $N_{\rm H,MYT}=5.0^{+2.8}_{-1.3}\times10^{24}$ cm$^{-2}$ and $N_{\rm H,borus}=2.7^{+0.4}_{-0.8}\times10^{24}$ cm$^{-2}$. Moreover, the covering factor found with the {\tiny BORUS} model is not consistent with the value of $C_f=0.5$ assumed in the {\tiny MYTORUS} one, and this might explain the difference in the column density estimate.\\
\indent In order to properly compare the two models, we construct a {\tiny BORUS} version of the {\tiny MYTORUS} decoupled model. We consider an out of line of sight reflector by setting the torus inclination to $\cos\theta=0.95$, which is the maximum value allowed by the {\tiny BORUS} model. This corresponds to an inclination angle of $\theta=18^\circ$, differently from the {\tiny MYTORUS} value $\theta=0$. The covering factor is fixed to the {\tiny MYTORUS} value $C_f=0.5$. As expected, the inclination discrepancy is not crucial \citep[see also][]{marchesi19} and we obtain $N_{\rm H,borus} = 5^{+3}_{-1} \times 10^{24}$ cm$^{-2}\simeq N_{\rm H,MYT}$. We stress that this model has been built with the sole purpose of comparing the column density of the torus for the {\tiny MYTORUS} to the one obtained with {\tiny BORUS}, since the goodness of fit is $\chi^2/{\rm dof}=944/902$, marginally worse than the model presented in Eq.~\ref{eq:borus_final}.\\
\indent However, this configuration is more realistic than the one with a covering factor of $C_v\sim0.9$, since the latter would imply that $\sim90\%$ of the sightline intercepts a Compton-thick column density. As a consequence, a Compton-thick state would be observed far more frequently. In fact, while this source has been observed several times, it has been caught in a Compton-thick state only once in 2011 by {\it Suzaku}. This would be possible if we were looking at this Seyfert galaxy with an exceptional, extremely polar, line of sight, whereas \cite{schmid03} estimated a $45^\circ$ angle for the inclination. Therefore a lower covering factor is likely a more realistic scenario for this source.

\subsection{Compton-thin absorber and Compton-thick reflector}

\indent Both models indicate that the absorbing material is Compton-thin, with column density ranging from $N_{\rm H,abs}\sim2\times10^{23}$ cm$^{-2}$ up to $N_{\rm H,abs}\sim4\times10^{23}$ cm$^{-2}$. This AGN was already caught in this state by one {\it Swift}-XRT snapshot in 2006 and by {\it XMM-Newton} in 2013. However, as shown by \cite{miniutti14}, the source is able to change from a relatively unobscured state ($N_{\rm H,abs}\sim2-4\times10^{22}$ cm$^{-2}$) up to a Compton-thick state.\\
\indent Previous analyses of ESO 323-G77 have hinted that low obscuration states ($N_{\rm H}\lesssim10^{23}$ cm$^{-2}$) might be caused by the presence of the obscuring torus, while higher obscuration states are likely due to absorption by cold intra-clump material located in the broad line region \citep{miniutti14,sanfrutos16}. However, given that we do not observe a change of state during the campaign analyzed in this work, but only moderate changes in the absorber column density $N_{\rm H,abs}$ we are not able to argue in favour or against this hypothesis.\\
\indent The unprecedented effective area of {\it NuSTAR} in the $E>10$ keV band allows us to properly study the reflection component of the X-ray spectrum of this source. In particular, both the {\tiny MYTORUS} and {\tiny BORUS} models clearly point to the presence of a Compton-thick reflector with $N_{\rm H}=5.0^{+2.8}_{-1.3}\times10^{24}$ cm$^{-2}$ or $N_{\rm H}=2.7^{+0.4}_{-0.8}\times10^{24}$, depending on the model. If the absorption is indeed produced by BLR clumps or intra-clump cold material, this result indicates that the constant Compton-thick reflector is located farther away from the central X-ray source, and it should be associated with the classic torus.

\subsection{Ionized absorbers}
\label{sec:iondisc}

Similar to the results presented in \cite{jimenez-bailon08}, \cite{miniutti14} and \cite{sanfrutos16}, our data shows the presence of two ionized absorbers.\\  
\indent We can estimate the location of these ionized absorbers using standard arguments. For instance, the maximum distance from the black hole can be estimated by considering that the size of the absorbing clump $R_{\rm clump}$ cannot be larger than the distance, i.e. $R_{\rm clump}=N_{\rm H}/n<r_{\rm max}$, where $n$ is the density of the clump \citep[e.g.,][]{crenshaw12,serafinelli21}. From the ionization parameter definition the maximum distance from the black hole can be written as
\begin{equation}
r_{\rm max}=\frac{L_{\rm ion}}{N_{\rm H}\xi}.
\label{eq:maxdist}
\end{equation}
\indent The first one, located in what we denote as Zone 1, is characterized by an ionization parameter of $\xi\sim250$ erg cm s$^{-1}$. The ionizing luminosity in the $E=13.6$ eV$-$ $13.6$ keV energy band is $L_{\rm ion}\simeq(2.6\pm0.2)\times10^{44}$ erg s$^{-1}$ and the column density is $N_{\rm H}\simeq3\times10^{23}$ cm$^{-2}$. Therefore, using Eq.~\ref{eq:maxdist}, we obtain $r_{{\rm max},1}=1.4^{+0.4}_{-0.9}$ pc.\\
\indent The second ionized absorber, located in Zone 2 is characterized by a larger ionization parameter, $\xi\simeq10^4$ erg cm s$^{-1}$. The average column density is given by $N_{\rm H}\simeq6\times10^{23}$ cm$^{-2}$. Therefore, using Eq.~\ref{eq:maxdist} we obtain a maximum distance of $r_{{\rm max},2}=1.0^{+0.9}_{-0.8}\times10^{-2}$ pc.\\
\indent We condider an Eddington ratio of $\log\lambda_{\rm Edd}=-0.56$ and a black hole mass of $M_{\rm BH}=2.5\times10^7$ $M_\odot$ \citep{wang07}, from which we can compute $\log L_{\rm bol}\simeq44.93$.This means, assuming that $L_{\rm bol}/L_{5100\AA}\sim10$ \citep[e.g.,][]{collin02}, that the optical luminosity is $\log L_{5100\AA}\simeq43.93$. We consider the relation between the broad line region size and the optical luminosity introduced by \cite{bentz09} $$\log R_{\rm BLR}({\rm light\;days})=-21.3+0.519\log L_{5100\AA}$$ and we obtain a broad line region radius of $R_{\rm BLR}\simeq0.02$ pc. We therefore obtain that the moderately ionized absorber in Zone 1 could be located outside the broad line region at $r_1\lesssim1.5$ pc, while the more ionized absorber in Zone 2 is likely co-spatial or within the BLR.\\ \indent In the scenario in which the cold absorber either co-spatial with one of the two ionized absorbers or sandwiched between them \citep{sanfrutos16}, the cold absorber would be located between the outer BLR, consistently with the model proposed by \cite{miniutti14}, and pc-scale distances. In the latter case, a possible scenario would be the presence of an inner thick reflecting ring surrounded by a thinner absorbing layer at pc-scale \citep[e.g.,][]{buchner19}. Recent mid-infrared results \citep{leftley21} found evidence of the presence of polar warm dust at a distance $r\gtrsim1.5$ pc, which is consistent with this scenario. The outer layer would also be clumpy, allowing for the observed long-term variability, a similar scenario to the one proposed for NGC 7479 by \cite{pizzetti22}.

\subsection{Ultra-fast outflow}

The velocity of the absorber in Zone 2 is $v\lesssim9000$ km s$^{-1}$, consistent with the values measured by \cite{jimenez-bailon08} and \cite{sanfrutos16} of $v\simeq2000$ km s$^{-1}$, in Epochs 1-4. However, in Epoch 5, we notice a moderately relativistic velocity $v\sim0.21c$, with a level of $\Delta\chi^2/{\rm dof}=11/1$. This is a tentative indication that we are observing an absorber outflowing at high velocity, a phenomenon that is commonly known as ultra-fast outflows (UFOs) and are fairly common ($\sim40\%$) in Seyfert galaxies and quasars \citep[e.g.,][]{pounds03,braito07,tombesi10,gofford13,nardini15,tombesi15,serafinelli19}. Moreover, UFOs are known to be extremely variable \citep[e.g.,][]{reeves14,matzeu17,braito18,braito22}, therefore it is not surprising that the UFO appears within a relatively short timescale in an AGN that never showed signs of its presence before. However, given its  modest ($\sim3\sigma$) detection here, further observations would be required to confirm the detection of the UFO feature or its variability.\\

\subsection{Coronal parameters}

\begin{figure}
\centering
\includegraphics[scale=0.35]{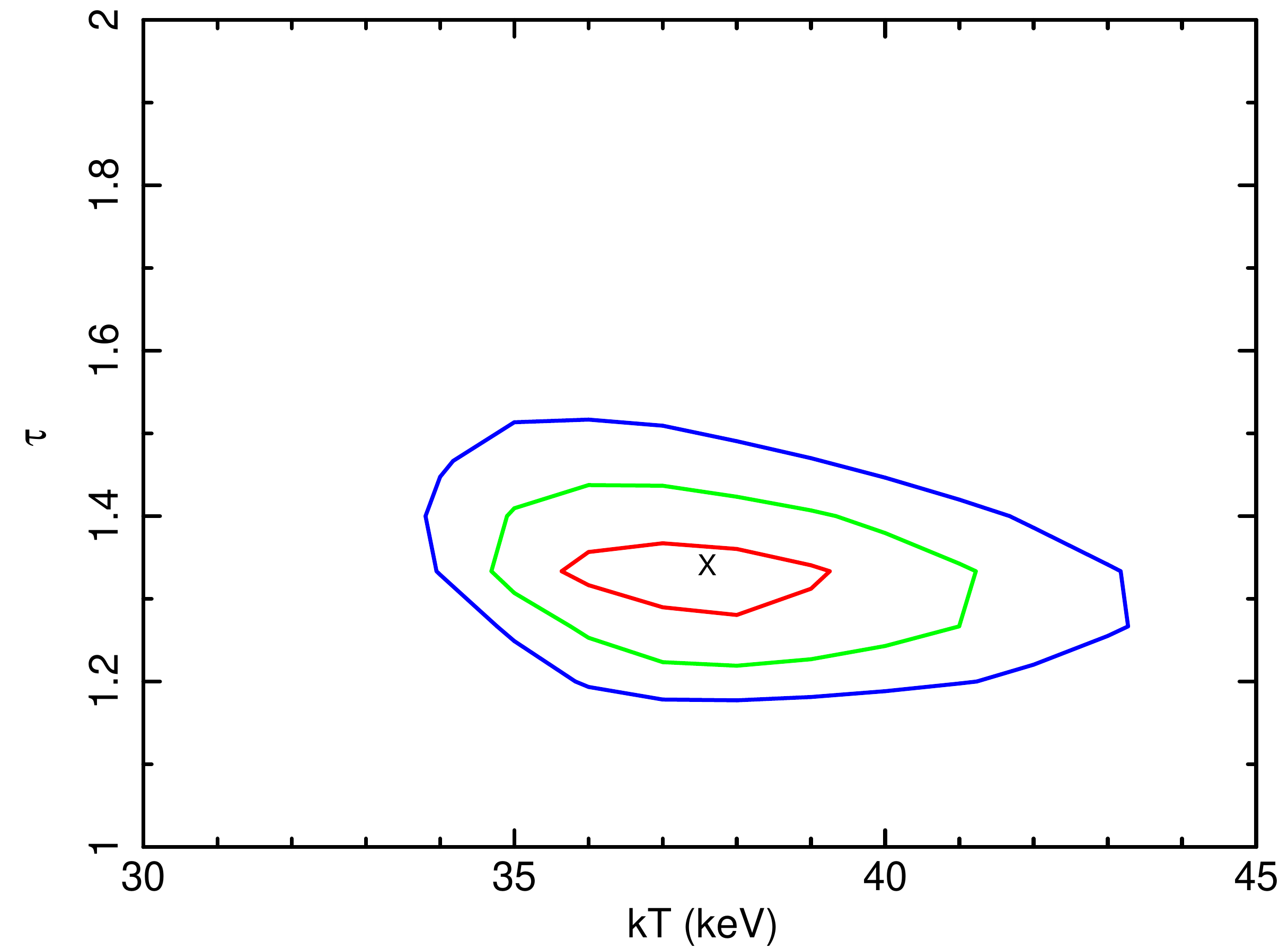}
\caption{Contour plot of the optical depth $\tau$ versus the coronal temperature $kT_e$, assuming a slab coronal geometry with the {\tiny COMPTT} Comptonization model. The red, green and blue lines represent $68\%$ ($1\sigma$), $95\%$ ($2\sigma$) and $99.7\%$ ($3\sigma$) confidence levels, respectively.}
\label{fig:contkTtau}
\end{figure}

The X-ray continuum is well known to be produced by inverse Compton on UV seed photons gaining energy by a very hot electron corona \citep[e.g.,][]{haardt91,haardt93}. The electron temperature therefore plays a crucial role in regulating the Comptonization of UV seed photons. Indeed, the main continuum breaks at the so-called cut-off energy $E_{\rm cut}$, which is tied to the temperature by the relation $E_{\rm cut}=2-3\,kT_e$, depending on the geometry of the corona \citep[e.g.,][]{petrucci01}.\\
\indent When the {\tiny COMPTT} model is adopted to model the continuum, assuming a slab geometry for the corona, in the {\tiny MYTORUS} model shown in Eq.~\ref{eq:mytcomptt}, we find that the temperature of the corona is $kT_e=38\pm2$ keV, with an optical depth $\tau=1.4\pm0.1$. The $\tau-\Gamma$ contour plot is shown in Fig.~\ref{fig:contkTtau}. Since the grids do not allow much larger values of $\tau$, the only way to study the spherical geometry is to use the {\tiny NTHCOMP} Comptonization continuum with the {\tiny BORUS} model (Eq.~\ref{eq:borus_nthcomp}), and we find a consistent temperature, although with larger errors, $kT_e=36^{+13}_{-8}$ keV.  We can estimate the optical depth using the following equation, valid for a spherical optically thick ($\tau>1$) corona \citep{zdziarski96}: $$\Gamma=\sqrt{\frac{9}{4}+\frac{511\;{\rm keV}}{kT\,\tau(1+\frac{\tau}{3})}}-\frac{1}{2}.$$ Using the best-fit values of the BORUS model, summarized in Table~\ref{tab:borustab}, we obtain $\tau\simeq2.8$.\\
\indent These are fairly standard values, as the coronal temperature is known to span from $kT\sim3$ keV up to $kT\sim450$ keV \citep[e.g.,][Serafinelli et al., in prep.]{matt15,tortosa18,tortosa22}. However, even though some authors have recently unveiled coronal temperatures in isolated obscured sources \citep[e.g.,][]{middei21} and samples of Seyfert 2 galaxies \citep[e.g.,][]{balokovic20}, they are not easily constrained, since they are often degenerate with the reflection spectrum cut-off.

\section{Summary and conclusions}
\label{sec:concl}

We have presented the spectral analysis of a campaign of 5 {\it NuSTAR} observations of the Seyfert 1.2 galaxy ESO 323-G77. We summarize our results in the following

\begin{itemize}
\item The source has been observed in a persistent obscured, but Compton-thin state, due to the presence of neutral obscuring material on the line of sight, with column density in the range $N_{\rm H}\sim2-4\times10^{23}$ cm$^{-2}$.
\item We find a Compton-thick reflector both modelling it with {\tiny MYTORUS} and {\tiny BORUS}. The two $N_{\rm H,refl}$ values are not consistent, but this result is dependent on the covering factor of the reflector, which is assumed as $C_f=0.5$ in {\tiny MYTORUS} and fitted ($C_f=0.90^{+0.02}_{-0.03}$) in {\tiny BORUS}. By fixing a more realistic $C_f=0.5$ in {\tiny BORUS}, the two results are consistent.
\item Two ionized absorbers are needed in our models, consistent with \cite{jimenez-bailon08}, \cite{miniutti14} and \cite{sanfrutos16}. The ionized absorber identified with Zone 1 is located at a distance of about $r_1\sim1.5$ pc from the black hole, most likely outside the broad line region, whose size is estimated as $R_{\rm BLR}\simeq0.02$ pc. The ionized absorber in Zone 2 instead is located at $r_2\simeq10^{-2}$ pc, either co-spatial or within the BLR.
\item Assuming that the cold absorber is either at the same distance of one of the two ionized absorbers, or at an intermediate one, its location can be placed between the outer BLR and at pc-scale distances. In the first case, this would be consistent with the model proposed by \cite{miniutti14}, consisting of cold absorbing intra-clump material in the BLR. In the second case, the most likely scenario is pc-scale Compton-thin absorbing material surrounding a Compton-thick reflector \citep{buchner19}, which is supported by recent mid-infrared detection of polar dust at $r\gtrsim1.5$ pc \citep{leftley21}.
\item The ionized absorber in Zone 2 is blueshifted  at Epoch 5, to the value $z_{\rm obs}\simeq-0.18$, which suggests an outflowing velocity of $v_{\rm out}\simeq0.2c$. 
\item The coronal temperature is constrained in both models, finding $kT_e\simeq37$ keV, both assuming a slab and a spherical corona. The optical depth is $\tau\simeq1.4$ when the slab coronal geometry is assumed, and $\tau\simeq2.8$ for a spherical corona.
\item We find hints of the possible presence of a relativistic reflection component from the accretion disk. However, this component contributes to $\lesssim10\%$ of the observed $2-10$ keV flux, and it mostly affects the $3-5$ keV energy band. Hence, the parameters of the disk reflection component are very difficult to constrain, and higher energy resolution data are needed to further study this feature.
\end{itemize}

\noindent The campaign was not able to observe any significant change of state (e.g., obscured to unobscured), as the source has undergone several times in the past \citep{miniutti14}. However, longer campaigns should be able to observe the source passing from obscured to unobscured or vice-versa, setting an upper limit to the obscurer location. Future high-resolution instruments such as the microcalorimeter Resolve on board XRISM \citep{xrism20} will be able to measure the properties of the absorbers with much more detail, particularly on their location and outflowing velocity. Moreover, future hard X-ray ($E=2-200$ keV) instruments such as the High Energy X-ray Probe \citep[HEX-P,][]{madsen18} will allow us to measure the reflection parameters with unprecedented accuracy.

\begin{acknowledgements}
      The authors thank the referee for useful comments that improved the quality of this paper. RS, VB, PS, ADR, and RDC acknowledge financial contribution from the agreements ASI-INAF n.2017-14-H.0 and n.I/037/12/0. This research has made use of data and software provided by the High Energy Astrophysics Science Archive Research Center (HEASARC), which is a service of the Astrophysics Science Division at NASA/GSFC and the High Energy Astrophysics Division of the Smithsonian Astrophysical Observatory. This research has made use of the {\it NuSTAR} Data Analysis Software (NUSTARDAS) jointly developed by the ASI Space Science Data Center (SSDC, Italy) and the California Institute of Technology (Caltech, USA). We acknowledge the use of public data from the {\it Swift} data archive.
\end{acknowledgements}

%
   \bibliographystyle{aa} 
   \bibliography{biblio} 
%

%
%

%
%
 




\end{document}